\documentclass[iop]{emulateapj}

\begin{document}


\title{Study of the Effect of nearest neighbors on Ferromagnetic to paramagnetic phase transition in 2D lattices by Monte Carlo algorithm}


\author{Saeed Rahmanian Koshkaki\altaffilmark{1,2} \\ \href{mailto:saeed.koshkaki@skolkovotech.ru}{saeed.koshkaki@skolkovotech.ru}}
\affil{Moscow Institute of Physics and Technology, 9 Institutskiy lane, Dolgoprudny city, Moscow Region 141700, Russia\altaffilmark{1}}

\affil{\\Skolkovo Institute of Science and Technology, 3 Nobel Street, Skolkovo 143025, Russia\altaffilmark{2}}




\begin{abstract}
In this work we try to use the Monte Carlo algorithm, metropolis, to study the behavior of 2D magnetic systems; honeycomb, hexagonal and square lattices. In this study we use Ising Model of magnetism, with considering only nearest neighbors in the Hamiltonian and with the Ferromagnetic correlation constant. The code was written in Payton. In this calculation, by comparing the results obtained for all lattices, we will show that the paramagnetic to ferromagnetic phase transition depends on the number of neighbor and it is a direct dependence.
\end{abstract}

\section{Introduction}

There are different types of magnetic materials, Ferromagnet, Paramagnetic, Diamagnetic  and Ferromagnetic. For example hard magnets are an example of ferromagnetic and superconductors are diamagnetic. Some magnetic structures could be either ferromagnetic  or paramagnetic, depend on the temperature and external magnetic field. By decreasing the temperature or increasing external magnetic field, we can see transition from paramagnetic to ferromagnet. The transition criteria (temperature, external magnetic field, etc.) depends on many factors, both macroscopic (size of the sample) or microscopic interactions (Spin-Orbit coupling, hopping integral, etc.). In this paper we try to study one of the microscopic factors, which is the number of nearest neighbor atoms. For this, we have selected three main 2D structures, honeycomb, square and hexagonal, respectively with 3, 4 and 6 nearest neighbor to each atom in the lattice. For honeycomb lattice we can consider graphene to visualize the system. In our study the strength of correlation between two neighboring atoms is same for all three lattices. In this study, we performed completely abstract calculation for all lattices that is made of atoms with net angular momentums ($µ_m$ = 1). It means, we try to show the difference between a honeycomb, hexagonal and square lattices phase transition from ferromagnetic to paramagnetic not to compute some quantities to predict the transition temperature. for doing this comparison, we calculated physical properties of system. In our study we used Ising model for a magnetic insulator, in fact we assumed that the systems we are studying has a none-zero gap on Fermi level.

The organization in this paper is in this way that, at first we will explain the calculation method in section 2, then in section 3 we will discuss the results and outcomes of calculations and finally we will give a short conclusion.

\section{METHODOLOGIES}


In Ising Model strong and energetically unfavorable intra-atomic Coulomb repulsion is ignores. This approach is related to the Heitler - London approximation in molecular science[4]. Hamiltonian in Ising model for location i-th on the lattice is defined as, 

\begin{equation}
H_i = -J \sum_{<ij>}^{N} \mu^i_z.\mu^j_z - \sum_{i}^{N} B^i_z.\mu^i_z ,
\end{equation}
In this Hamiltonian $\mu$ stands for the magnetic moment on position i-th(with values +1 and -1), and J indicates the effective interaction between two magnetic moments. $J>0$ stands for ferromagnetic order and $J<0$ is for anti-ferromagnetic order. Also the magnetic momentum(spin) could be only directed in $z$ direction with states of up $\mu^j_z=1$ and down $\mu^j_z=-1$ magnetic momentum. This model is known as Ising model, which was proposed for the first time by Wilhelm Lenz in 1920.  $B^i_z$ is the local magnetic in the i-th location.$<ij>$ indicate the sum over all nearest neighbors, in fact the summation must be written in the form $\sum_{i}^{N}\sum_{j}^{n} S_i.S_j$. Here we assume that The value of J for the nearest neighbor for all three lattices is the same so we can compare the effect of nearest neighbors number on transition temperature. Since then, in all calculations we set all constant to one, including Plank constant, Boltzmann constant and Bohr magneton.

In this study because we are interested in ferromagnetic state, we set $J>0$. Since the purpose of this study is to compare the phase transition in 3 different lattices, the temperature is a multiple of J, T = $\alpha$J, where $\alpha$ is constant. 






\subsection{PHYSICAL QUANTITIES} 


In this study, like many other studies in Ising model, we calculated 3 main properties for all three systems; Susceptibility, Magnetization and Heat Capacity. Also in a separate section we will show the effect of external magnetic field ($B^i_z$) on the transition temperature.

We defined the net magnetization by following equation,
\begin{equation}
<M> = \frac{\sum_{i}^{N} \mu^i_z exp(-\frac{H_i}{T})}{\sum_{i}^{N} exp(-\frac{H_i}{T})}
\end{equation}

Where, $H_i$ is the energy of i-th position and N is the total number of atoms in studied lattice. Thermodynamically we can define susceptibility as,
\begin{equation}
\chi = \frac{\partial <M>}{\partial (\frac{1}{T})}
\end{equation}

Also heat capacity by,
\begin{equation}
\chi = \frac{\partial <E>}{\partial T}
\end{equation}

In the equation (4), $<E>$ has the following form,
\begin{equation}
<E> = \frac{\sum_{i}^{N} H_i exp(-\frac{H_i}{T})}{\sum_{i}^{N} exp(-\frac{H_i}{T})}
\end{equation}
which is the average energy of system. By doing some simple algebra, that you can find in any advanced Statistical Mechanics book, we reach the following equations for susceptibility and heat capacity,
\begin{equation}
\chi = \frac{1}{T}(<E^2>-<E>^2)
\end{equation}

\begin{equation}
\chi = \frac{1}{T^2}(<M^2>-<M>^2)
\end{equation}

\subsection{MODELING THE SYSTEM} 
For modeling the system we made a matrix for each structure with same number of atoms for each. In our system we set the boundary to be ferromagnetic, with stronger coupling coefficient (J) than in our lattice, so it will not be affected by changing the temperature of modeled lattice. In all the cases, the macroscopic lattice is square for consistency of the calculation, while for example, honeycomb and hexagonal could have macroscopic hexagonal shape. In figure one you can see a sample of our lattices,

\begin{figure}
\includegraphics[scale=0.29]{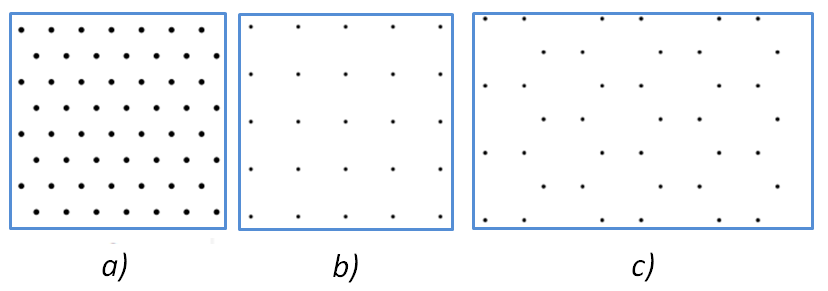}
\caption{Crystal structure of three lattices: a)Hexagonal, b)Square, c)Honeycomb 
.}
\end{figure}

In Square lattice, the boundary is a simple linear atomic chain, for honeycomb we have armchair and zigzag boundary, and for hexagonal the boundary is either linear or zigzag.
 
The algorithm we used is the Monte Carlo algorithm, specifically metropolis, to simulate the system behavior in each case study.
\section{RESULTS}
In this section we represent our result for two different part of study. Firstly, we will discuss and compare the transition temperature of physical properties of all three systems when external field, $B^i_z$ in equation (1), is zero. Secondly we will turn on the on-site effective magnetic field and will show the dependence of phase transition on strength of external magnetic field.
\subsection{PHYSICAL PROPERTIES FOR $B_i=0$} 
When the external magnetic field is zero the only interaction which is responsible for magnetic order in the system is coupling between magnetic moments. In high temperature (higher than transition temperature) the kinetic energy from temperature is an obstacle to creation of order(ferromagnetic) in disorder magnetic systems(paramagnetic), as we reduce the temperature the reduction in kinetic energy will happen. Bellow the transition temperature, the coupling between magnetic moments become significant that it can overcome the kinetic energy of the system and the order will start to be created in some domains of crystal. In Figure 2 you can easily see that, as we reduce the temperature the net magnetization after some temperature become non-zero.

\begin{figure}
\includegraphics[scale=0.21]{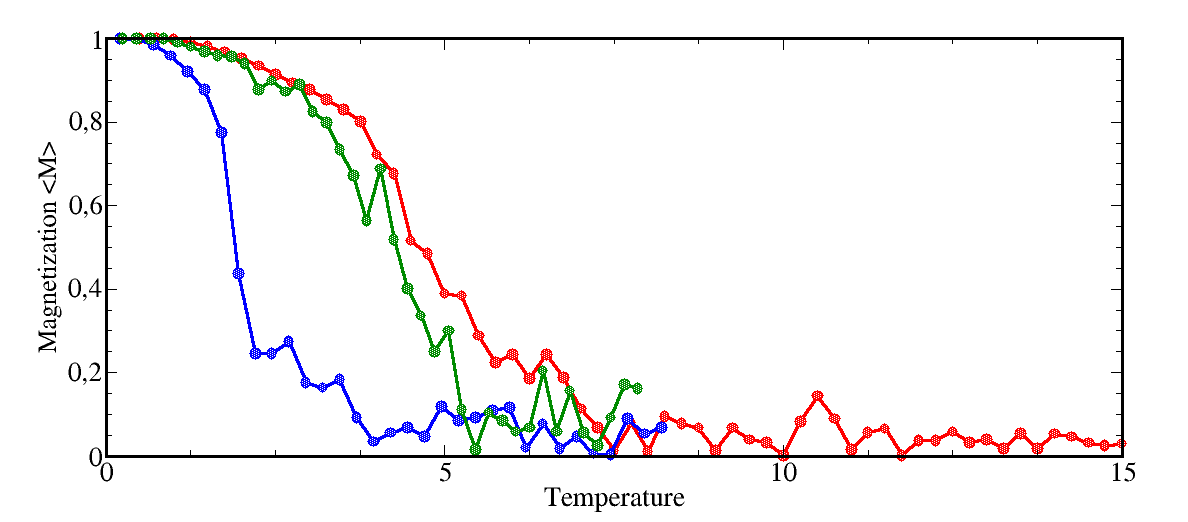}
\caption{Magnetization of each lattice calculated by equation(2). Red for hexagonal, green for square lattice and blue for honeycomb lattice.}
\end{figure}

 If we look carefully we can see something interesting at temperature where the creation of net magnetization happens (transition temperature). In fact, this figure shows that he highest transition temperature belongs to the hexagonal lattice, and the lowest one to honeycomb lattice. This is happening because of the number of nearest neighbors, which is 6 for hexagonal and 3 for honeycomb. The higher number of nearest neighbor the stronger magnetic moment’s interactions, so in higher temperature the system can overcome the kinetic energy due to the temperature. This is exactly what we expect from mean field approximation of Ising model[5]. According to mean field model, transition temperature is defined as $T_c=mA$, where A is some constants and m is the number of neighbors. Be aware that in figure 1, the numbers on x axis is just to simplify the comparison not to predict the value of temperature; this will be the same for all later figures.

Another physical property which we calculated for the three systems, is specific heat and susceptibility, which are depicted in figure 3 and 4.

\begin{figure}
\includegraphics[scale=0.21]{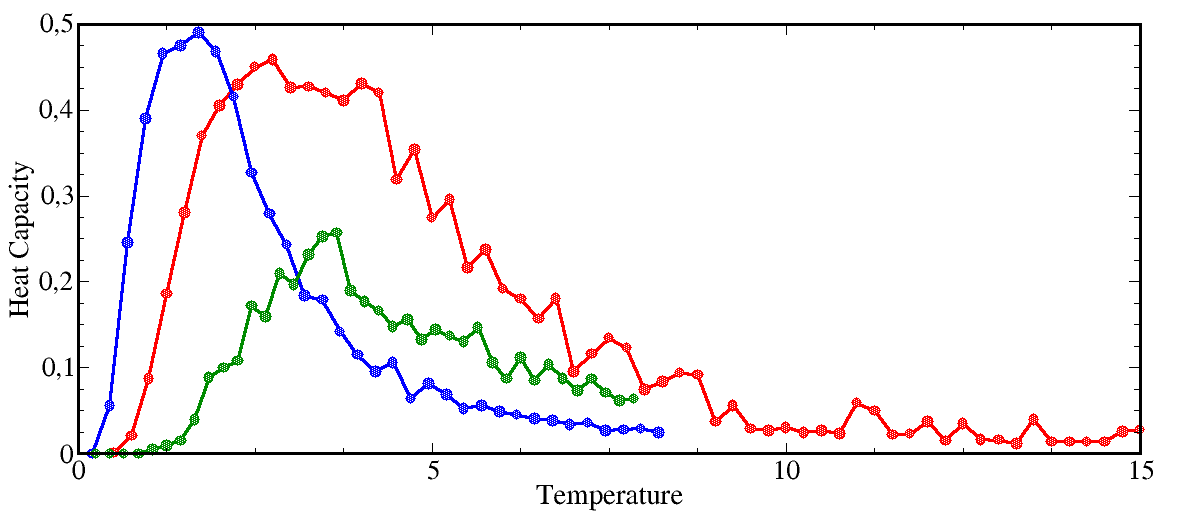}
\caption{Heat capacity calculated for all lattices. Red for hexagonal, green for square lattice and blue for honeycomb lattice.}
\end{figure}

\begin{figure}
\includegraphics[scale=0.21]{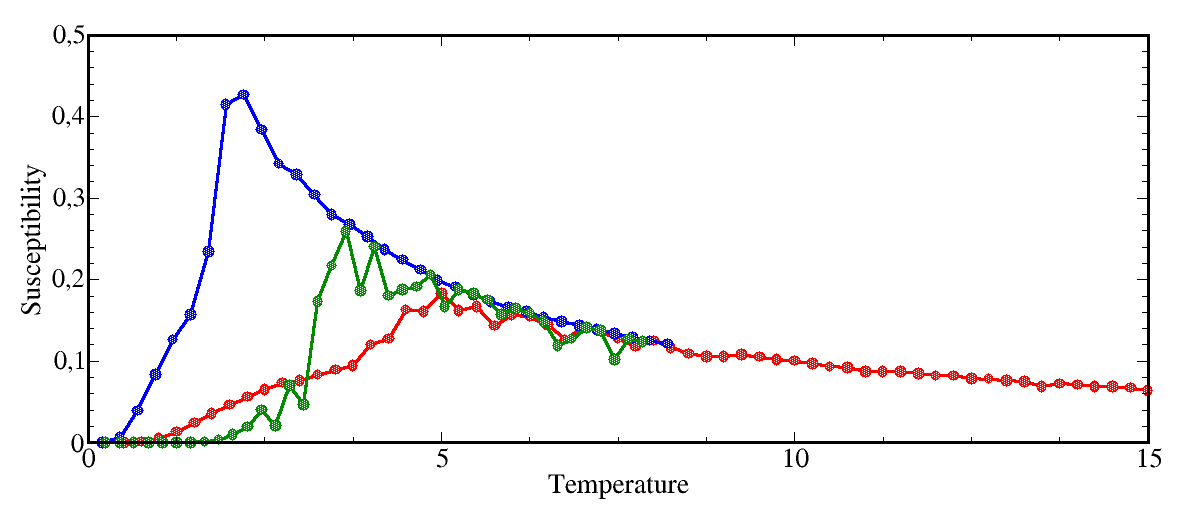}
\caption{Susceptibility calculated for all lattices. Red for hexagonal, green for square lattice and blue for honeycomb lattice.}
\end{figure}

For the phase transition from ferromagnetic to paramagnetic, on transition temperature, the values of heat capacity and susceptibility will have a peak as shown in figures 3 and 4. Again by increasing number of nearest neighbors we have higher value of transition temperature (the transition happens when heat capacity and Susceptibility reach their maximum ). Another important fact which could be infer from figure 3 and 4 is that by increasing the number of neighbors, the peak will be higher. The only disagreement is in square lattice heat capacity peak value, which is lower than others, and it could be from numerical errors. Also for Susceptibility when the temperature is high enough the effect of neighbors become negligible and all three lattice reach the same value of susceptibility and its change almost linearly by temperature (the plots in the figure 4 above T = 5 )

Alongside these physical properties we calculated the energy of systems, shown in figure 5. This figure shows the lower energy in system happens when there is magnetic order.

\begin{figure}
\includegraphics[scale=0.21]{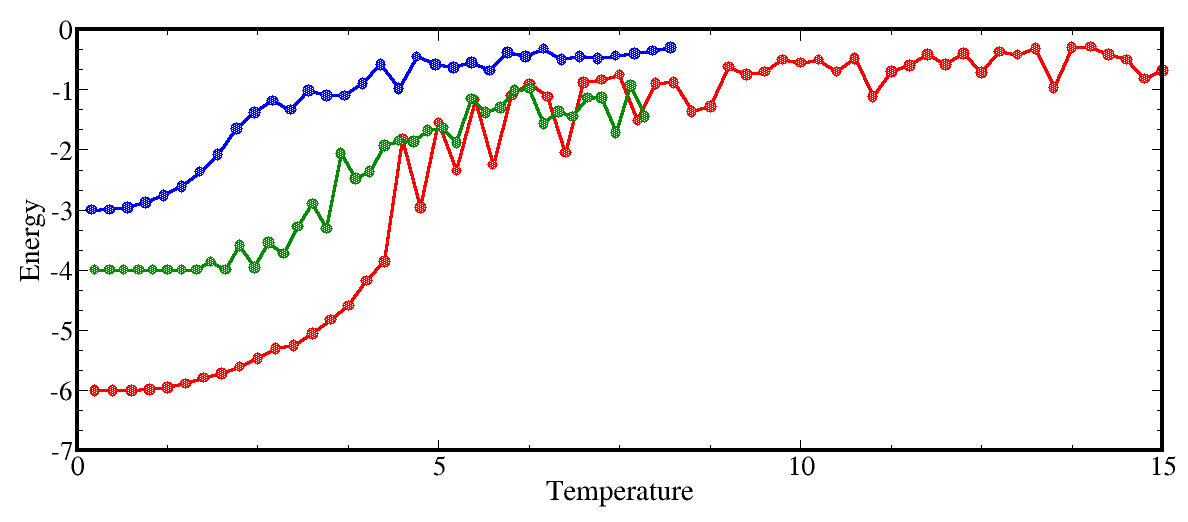}
\caption{Average energy of each system by equation (5). Red for hexagonal, green for square lattice and blue for honeycomb lattice.}
\end{figure}

\subsection{TRANSITION TEMPERATURE FOR NON-ZERO $B_i$} 
In this part we will show the effect of external magnetic field on transition temperature. First of all we should explain some about the way we calculated the transition temperature. Due the numerically errors in calculations, to make our results more reliable, we will report two temperature, which the transition temperature will have a value between them. We define total magnetic moment ($M_t$) as follow, 
\begin{equation}
<M_t> = \frac{\sum_{i}^{N} \mu^i_z }{N}
\end{equation}
This function has values between 0 and 1. In reported temperature. first temperature is the temperature at which the total magnetic moment reach 0.9, and the second reported number is the temperature at which $M_t$ will have values less than 0.2. The plot figure for this temperatures could be found in figure 6. By looking at this plotting we can see the increase in transition temperature by increase the magnetic field. The value of temperature at $M_t = 0.9$, has linear dependence to the value of external field, but for when $M_t<0.2$, the temperature has $a^ B$ dependency, where $a$ is a positive number and B is magnetic field. This is happen due to the coupling effect, when $M_t=0.9$ the coupling effect is high so it is harder to break the coupling between moments, but when $M_t=0.2$ the weak coupling cannot resist against the kinetic energy so the breaking is easier. The increase in transition temperature by increasing magnetic field could be easily deduced from the second term in equation (1), where this term tries to decrease the effect of kinetic energy.

\begin{figure}
\includegraphics[scale=0.4]{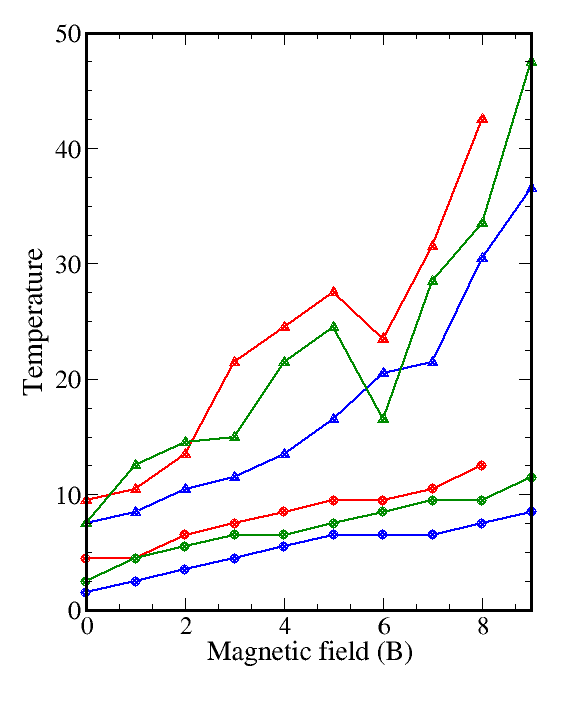}
\caption{Transition temperature versus external magnetic field. Red for hexagonal, green for square lattice and blue for honeycomb lattice. Circles are for when $M_t = 0.9$ and triange for when $M_t < 0.2$}
\end{figure}

\section{CONCLUSION}
By using Ising model we have been able to compare the effect of number of nearest neighbors on transition temperature form paramagnetic to ferromagnetic. According to this research, we predict that the more nearest neighbors in lattice will end up to the higher transition temperature, as has been shown for magnetization, heat capacity and susceptibility of all three lattices. Also in susceptibility, we saw that peak at transition temperature has higher value for bigger number of nearest neighbors. Finally by turning on the external magnetic field we showed that in the presence of external magnetic field, the transition temperature will increase and the dependence is exponential.\\

\section{ACKNOWLEDGMENTS}

This work is supported by the Russian Science Foundation grant in Moscow Institute of Physics and Technology and scholarship from Skolkovo Institute of Science and Technology. 

\acknowledgments

\end{document}